\documentclass[12pt]{article}

\usepackage{etex}
\usepackage{amsmath,amssymb,amsthm,amscd,amsbsy,array}

\usepackage{color}
\usepackage[colorlinks=true, pdfstartview=FitV, linkcolor=blue, citecolor=blue, urlcolor=blue]{hyperref}

\let\ssection=\section
\renewcommand{\section}{\setcounter{equation}{0}\ssection}

\newcommand{\bR}{{\bf R}}

\newcommand{\Aut}{\mathrm{Aut}}
\newcommand{\bb}{\mathbf{b}}
\newcommand{\Barg}{\mathrm{Barg}}

\newcommand{\bc}{\mathbf{c}}
\newcommand{\Chr}{\mathrm{Chr}}

\renewcommand{\d}{{n}}

\newcommand{\Conf}{\mathrm{Conf}}

\newcommand{\hdelta}{{\widehat{\delta}}}
\newcommand{\hDelta}{{\widehat{\Delta}}}
\newcommand{\Diff}{\mathrm{Diff}}

\newcommand{\GL}{\mathrm{GL}}

\newcommand{\grad}{{\mathbf{grad}}}
\newcommand{\rg}{\mathrm{g}}
\newcommand{\hrg}{{\widehat{\rg}}}

\newcommand{\hh}{h} 
\newcommand{\hath}{\widehat{h}} 

\newcommand{\cM}{\mathcal{M}}

\newcommand{\bomega}{\boldsymbol{\omega}}
\newcommand{\bOmega}{\boldsymbol{\Omega}}
\newcommand{\homega}{{\widehat{\omega}}}
\newcommand{\hbomega}{{\widehat{\bomega}}}

\newcommand{\hbOmega}{{\widehat{\bOmega}}}
\newcommand{\hpartial}{{\hat{\partial}}}
\newcommand{\PSL}{\mathrm{PSL}}
\newcommand{\PGL}{\mathrm{PGL}}

\newcommand{\tPsi}{\widetilde{\Psi}}
\newcommand{\tpsi}{\widetilde{\psi}}

\newcommand{\htPsi}{(\tPsi)^{\widehat{\ }}}
\newcommand{\cQ}{{\mathcal{Q}}}

\newcommand{\hm}{{\widehat{m}}}
\newcommand{\tn}{N}
\newcommand{\hnabla}{{\widehat{\nabla}}}

\newcommand{\Ric}{\mathrm{Ric}}

\newcommand{\cS}{{\mathcal{S}}}

\newcommand{\SN}{Schr\"odinger-Newton\ }
\newcommand{\rSN}{\mathrm{SN}}

\newcommand{\rO}{\mathrm{O}}

\newcommand{\hs}{{\hat{s}}}
\newcommand{\stars}{{s^*}}
\newcommand{\Sch}{\mathrm{Sch}}

\newcommand{\wht}{{\hat{t}}}
\newcommand{\start}{{t^*}}
\newcommand{\htheta}{{\widehat{\theta}}}

\newcommand{\hU}{{\widehat{U}}}

\newcommand{\vol}{\mathrm{vol}}
\newcommand{\Vol}{\mathrm{Vol}}
\newcommand{\bx}{\mathbf{x}}
\newcommand{\hx}{{\hat{x}}}
\newcommand{\hbx}{{\hat{\bx}}}
\newcommand{\starbx}{{\bx^*}}
\newcommand{\hxi}{{\widehat{\xi}}}
\newcommand{\by}{\mathbf{y}}

\newcommand{\half}{\frac{1}{2}}

\newcommand{\la}{{\langle}}
\newcommand{\ra}{{\rangle}}

\newcommand{\bigbox}[1]{\fbox{%
\rule[-20pt]{0pt}{45pt}$\;\;\displaystyle{#1}\;\;$}%
}
\newcommand{\medbox}[1]{\fbox{%
\rule[-15pt]{0pt}{35pt}$\;\;\displaystyle{#1}\;\;$}%
}

\begin{document}

\baselineskip=15.5pt

\oddsidemargin .1truein
\newtheorem{thm}{Theorem}[section]
\newtheorem{lem}[thm]{Lemma}
\newtheorem{cor}[thm]{Corollary}
\newtheorem{pro}[thm]{Proposition}
\newtheorem{ex}[thm]{Example}
\newtheorem{rmk}[thm]{Remark}
\newtheorem{defi}[thm]{Definition}

\title{
On the \SN equation\\ and its symmetries: a geometric view\\[6pt]
}

\author{
C. DUVAL\footnote{mailto: duval-at-cpt.univ-mrs.fr}
\quad\hbox{and}\quad
S. LAZZARINI\footnote{mailto: lazzarini-at-cpt.univ-mrs.fr}\\[5mm]
Centre de Physique Th\'eorique,\\
{Aix Marseille Universit\'e \& Universit\'e de Toulon \& CNRS UMR 7332,}\\
{Case 907, 13288 Marseille, France.}
}

\date{June 22, 2015}

\maketitle

\begin{abstract}
The \SN (SN) equation is recast on purely geometrical grounds, namely in terms of Bargmann structures over $(\d+1)$-dimensional Newton-Cartan (NC) spacetimes. Its maximal group of invariance, which we call the SN group, is determined as the group of conformal Bargmann automorphisms that preserve the coupled Schr\"odinger and NC gravitational field equations. Canonical unitary representations of the SN group are worked out, helping us recover, in particular, a very specific occurrence of dilations with dynamical exponent $z=(\d+2)/3$.   
\end{abstract}

\newpage

\tableofcontents

\newpage

\section{Introduction}

The \SN (SN) equation for a quantum non-relativistic particle of mass $m$ in Euclidean space of dimension $\d>2$ reads
\begin{equation}
i\hbar\frac{\partial\psi}{\partial{t}}(\bx,t)
=
\left(
-\frac{\hbar^2}{2m}\Delta_{\bR^\d}
-4\pi G m^2 C_\d\!\int_{\bR^\d\setminus\{\bx\}}
{
\!\frac{\vert\psi(\by,t)\vert^2}{\Vert\bx-\by\Vert^{\d-2}}\,dy^1\cdots{}dy^\d
}
\right)
\psi(\bx,t)
\label{SNEq}
\end{equation}
where $C_\d=\Gamma(\d/2)/(2\pi^{\d/2}(\d-2))$. It has originally been introduced as a special case, $N=1$, of the Schr\"odinger equation for a $N$-particle system in mutual gravitational interaction~\cite{Dio}. The non-local and non-linear differential equation (\ref{SNEq}) is in fact obtained, as explained below, by a specific coupling between the Schr\"odinger equation and the non-relativistic gravitational field equation. In the Schr\"o\-dinger equation,
\begin{equation}
i\hbar\frac{\partial\psi}{\partial{t}}(\bx,t)
=
\left(
-\frac{\hbar^2}{2m}\Delta_{\bR^\d}
+V(\bx,t)
\right)
\psi(\bx,t)
\label{SchEq}
\end{equation}
the  potential energy $V(\bx,t)=m\,U(\bx,t)$ is identified to the Newtonian gravitational potential, solution of the Poisson equation\footnote{
The Green function of the $\d$-dimensional Poisson equation (vanishing asymptotically) is given by $G(\bx,\by)=-C_\d/\Vert\bx-\by\Vert^{\d-2}$, and reduces to the familiar expression $G(\bx,\by)=-1/(4\pi\Vert\bx-\by\Vert)$ for $\d=3$.
}
\begin{equation}
\Delta_{\bR^\d}U(\bx,t)=4\pi G \varrho(\bx,t)
\label{Poisson}
\end{equation}
where $G$ stands for Newton's constant; the mass density of the sources is given here in terms of the quantum probability density, namely by
\begin{equation}
\varrho(\bx,t)=m\,\vert\psi(\bx,t)\vert^2
\label{varrho}
\end{equation}
where wave-function, $\psi$, is tacitly assumed to be normalized, $\int_{\bR^\d}{\!\vert\psi(\bx,t)\vert^2\,dx^1\cdots{}dx^\d}=1$.

Let us emphasize that, apart from its own mathematical interest as a highly non-trivial system (\ref{SchEq}), (\ref{Poisson}) and (\ref{varrho}) of coupled PDE, the \SN equation may, more significantly, be considered as instrumental to support Penrose's proposal for the ``Gravitization of Quantum Mechanics'' \cite{Pen1,Pen2}. A special class of stationary solutions can be found in, e.g., \cite{MPT,TM}. The \SN equation is also advocated for a dynamical interpretation of the process of reduction of wave packets in quantum mechanics, leading to a radical modification of the spreading of wave packets around and above a critical mass of the system \cite{GG,GG2}. See nevertheless \cite{BGDB} where the limitations to the quantum mechanical description of physical phenomena by the \SN equation are lucidly examined. This is however not the route that we are about to follow. 

\goodbreak

We will rather be interested in the \textit{symmetries} of this self-coupled quantum/classical system where the Schr\"odinger equation represents the quantum edge while the Newton field equation is granted a purely classical status within this coupled system. 

We will therefore aim at determining the maximal symmetry group of the SN equation. To some extent, this will be done in the wake and the spirit of earlier work of Niederer \cite{Nie} and Hagen \cite{Hag} related to the maximal symmetries of the (free) Schr\"odinger equation, after the discovery by S. Lie of those of the heat equation.\footnote{The so-called ``Schr\"odinger group'' has actually been discovered in a classical context by Jacobi in 1842; see, e.g., \cite{DH} for a review.} In doing so, we will find it worthwhile to generalize this coupled system (\ref{SchEq}), (\ref{Poisson}) and (\ref{varrho}), of partial differential equations so as to include non-inertial forces in ad\-dition to the Newtonian gravitational force, as well as a possibly curved Riemannian spatial background. This will be best achieved in geometrical terms by lifting this system of PDE to the Bargmann extended spacetime over the initially chosen Newton-Cartan (NC) spacetime; see \cite{DBKP} and references therein for an introduction to the subject.  The inverse problem of lifting a NC structure to what we call now a Bargmann structure has been pioneered by Eisenhart~\cite{Eis}. Since then, the formalism has been systematically utilized in various contexts ranging (highly non exhaustively) from the analysis of the fundamental interplay between the NC and (conformal) Lorentz geo\-metries \cite{DBKP,DGH,DL} to the study of hidden symmetries in quantum field theory \cite{JP,DHP1,DHP4} and classical Hamiltonian dynamics systems \cite{Cari,CGvHHZ}.

The main results of this article consist in (i) the generalization of the SN equation~(\ref{SNEq}) as given by (\ref{Yamabe}), (\ref{Mass}) and (\ref{NewtonSch}) by means of conformal Bargmann structures, and (ii) the determination of its maximal group of invariance, coined the \SN group (\ref{SN}). This vantage point enables us to treat 
in geometrical terms the rather complicated partial differential equation (\ref{SNEq}) and its natural generalization at one stroke. It also helps us produce the canonical unitary representations (\ref{rhodneq4}) and (\ref{rhod=4}) of the SN group on the set of solutions of the SN equation. As a special outcome, we recover  the representation of the specific dilation group with dynamical exponent $z=5/3$ discovered originally in~\cite{GG} in the case $n=3$. Let us stress that in the case of spatial dimension $\d=4$, the SN group gains one more dimension (time inversions are hence recovered), and is isomorphic to the well-known Schr\"odinger group (\ref{SN=Sch}) of $6$-dimensional Bargmann structures.

\medskip

The article is organized as follows.

Section \ref{BargmannSection} is devoted to a survey of Bargmann structures over nonrelativistic Newton-Cartan spacetime on which the rest of the article heavily relies.

Then, Section \ref{SNSection} establishes the geometric transcription of the \SN equation as a set of PDE on Bargmann extended spacetime which we call the generalized SN equations. Let us emphasize the central r\^ole granted to the Yamabe operator (or conformally-invariant Laplace-Beltrami operator). The SN equation (\ref{SNEq}) is duly recovered as a special case of the latter. 

\goodbreak

We introduce in Section \ref{SNSymmetries} the maximal group of symmetries of the (generalized) \SN equations, by taking advantage of the maximality of the conformal group as a group of symmetry of the Yamabe operator. We highlight, \textit{en passant}, the new expression (\ref{SchwarzianBis}) relating the Schwarzian derivative to the conformal transformation of the Ricci tensor. It is shown that mass-dilation occurs naturally in this context, leading to the fundamental constraint (\ref{dilationConstraint}) between the conformal dilation factor, $\lambda$, and the dilation factor, $\nu$, of the Bargmann fundamental vector field.

In Section \ref{SNgroupSection}, we define in full generality the so-called ``SN group'' of symmetry of our \SN system of equations. Explicit representations of this group are found in some particular instances, namely of the form (\ref{SNinGL}) for spatially flat Bargmann structures. The special case $\d=4$ is investigated along the same lines; see Footnote \ref{footnoteMaxwell} for a comment about this exceptional spatial dimension. We end this section by the quantum representations of these groups on the set of solutions, $\psi$, of the SN equation.

At last, Section \ref{ConclusionSection} help us summarize the main results of our work and announce new directions which constitute work in progress along these lines.



\section{Bargmann structures: a compendium}\label{BargmannSection}

\subsection{Main definitions}

A Bargmann structure is a triple $(M,\rg,\xi)$ where $M$ is a smooth connected and orientable manifold of dimension $\tn=\d+2$ with $\d>0$, endowed with a Lorentz metric $\rg$, and a complete, nowhere vanishing vector field $\xi$ which is light-like , $\rg(\xi,\xi)=0$, and covariantly constant with respect to the Levi-Civita connection, $\nabla\xi=0$ \cite{DBKP}. Thus $\xi$ can  be viewed as the fundamental vector field of a free action on $M$ of a $1$-dimensional Lie group, say $(\bR,+)$. The quotient $\cM=M/(\bR\xi)$ has been shown to be canonically endowed with a Newton-Cartan (NC) structure;\footnote{A NC structure is defined as a quadruple $(\cM,\hh,\vartheta,\nabla^\cM)$ where $\cM$ is a smooth $(\d+1)$-dimensional manifold, and $\hh$~a twice-contravariant tensor field whose kernel is generated by the nowhere vanishing closed $1$-form~$\vartheta$; moreover the torsionfree affine connection $\nabla^\cM$ parallel-transports $\hh$ and $\vartheta$. Let us recall that NC structures \cite{Car,Tra,Kun} (see also \cite{DH} and references therein) arise effectively from Bargmannian ones: if $\pi:M\to{}\cM$ is the submersion associated with a given Bargmann structure, then $\hh=\pi_*\rg^{-1}$, also $\theta=\pi^*\vartheta$, and $\nabla^\cM$ is the projection on $\cM$ of the Levi-Civita connection $\nabla$ of $(M,\rg)$.} it will be interpreted as non-relativistic spacetime. The quotient $T=M/\xi^\perp$ is endowed with the structure of a smooth ($1$-dimensional) manifold (either topology $T\cong\bR$ or $T\cong{}S^1$ is envisageable, see \cite{Duv1,DH}); we will denote by $\tau:\cM\to{}T$ the surjection over the \textit{time-axis} $T$.


The most general Bargmann structure is given locally by the pair\footnote{In the chosen coordinate system, $(x^1,\ldots,x^\d,t,s)$, the $\d$-uple $x=(x^1,\ldots,x^\d)$ provides coordinates on the intrinsically defined space at time $t$, denoted by $\Sigma_t$, while $s$ is a fiberwise coordinate of $\pi:M\to{}\cM$.}
\begin{equation}
\rg=\rg_{\Sigma_t}+dt\otimes\omega+\omega\otimes{}dt
\qquad
\&
\qquad
\xi=\frac{\partial}{\partial s}
\label{Bargmann}
\end{equation}
where 
\begin{equation}
\rg_{\Sigma_t}=\rg_{ij}(x,t)dx^i\otimes{}dx^j
\label{gSigma}
\end{equation}
is a preferred Riemannian metric on each time-slice $\Sigma_t=\tau^{-1}(\{t\})$ of $\cM$, and where
\begin{equation}
\omega=\omega_i(x,t)dx^i-U(x,t)dt+ds
\label{omega}
\end{equation}
is a connection form on the principal $(\bR,+)$-bundle $\pi:M\to{}\cM$.
The metric $\rg$ in~(\ref{Bargmann}) is known as a \textit{Brinkmann metric} \cite{Bri} or a generalized pp-wave \cite{EK}. The spacetime function~$U$ (for example, the profile of the gravitational wave whose wave-vector $\xi$ is null and parallel) is interpreted in the present context as the Newtonian gravitational potential on NC spacetime, $\cM$.

The $1$-form $\theta=\rg(\xi)$ associated with $\xi$ is covariantly constant, hence closed, and verifies $\theta(\xi)=0$; it thus descends to the time-axis $T$
as the ``clock'' of the structure, locally $\theta=dt$ (with some ruthless abuse of~notation). 

\bigskip
$\bullet$
Let us recall that the \textit{Bargmann group} (see, e.g., \cite{DBKP}), i.e., the group of strict automorphisms of a Bargmann structure, is defined by
\begin{equation}
\Barg(M,\rg,\xi)=\{\Phi\in\Diff(M)\,\strut\vert\,\Phi^*\rg=\rg,\Phi^*\xi=\xi\}
\label{Barg}
\end{equation}
where $\Phi^*$ denotes the pull-back operation by the diffeomorphism $\Phi$ of $M$.

The \textit{Schr\"odinger group} (see, e.g., \cite{DGH}) is the group of conformal auto\-morphisms of a Bargmann structure, namely
\begin{equation}
\Sch(M,\rg,\xi)=\{\Phi\in\Diff(M)\,\strut\vert\,\Phi^*\rg=\lambda_\Phi\rg,\Phi^*\xi=\xi\}
\label{Sch}
\end{equation}
where $\lambda_\Phi\in{}C^\infty(M,\bR^*_+)$.

\bigskip
$\bullet$
The canonical \textit{flat Bargmann structure} on $\bR^{\d+2}$ is given by\footnote{The group (\ref{Barg}) of automorphisms of the flat Bargmann structure (\ref{FlatBargmann}) is the group $\Barg(\bR^{\d+1,1})$, coined \textit{Bargmann group} by Souriau \cite{JMS}, generated by the transformations (\ref{hbx}), (\ref{wht}) and (\ref{hs}) where $d=g=1$ and $f=0$; see also the early reference \cite{GP}. If~$\d>2$, it is, up to equivalence, the unique $(\bR,+)$-central extension of the Galilei group of flat $(\d+1)$-dimensional NC spacetime as first shown by Bargmann \cite{Bar}. The Schr\"odinger group, $\Sch(\bR^{\d+1,1})$, i.e., the maximal symmetry group of the free Schr\"odinger equation \cite{Jac,Nie,Hag}, is again generated by the transformations (\ref{hbx}), (\ref{wht}) and (\ref{hs}) where $\nu=dg-ef=1$. The \textit{Schr\"odinger group} is, if $\d>2$, the unique $(\bR,+)$-central extension of the (centerless) Schr\"odinger group of flat $(\d+1)$-dimensional NC spacetime.}
\begin{equation}
\rg_0=\delta_{ij}\,dx^i\otimes{}dx^j+dt\otimes{}ds+ds\otimes{}dt
\qquad
\&
\qquad
\xi=\frac{\partial}{\partial s}.
\label{FlatBargmann}
\end{equation}

\subsection{Newton-Cartan gravitational field equations}

Let us recall, at this stage, that the NC gravitational field equations have been geometrical\-ly re\-formulated as \cite{DBKP}
\begin{equation}
\Ric(\rg)=4\pi{}G \varrho\,\theta\otimes\theta
\label{NewtonFieldEqs}
\end{equation}
where $\theta$ is the clock of the Bargmann structure, and $\varrho$ is the mass-density of the sources. We note that this implies that the scalar curvature vanishes identically, $R(\rg)=0$. 

\goodbreak

Let us mention that for the already interesting example of spatially-flat Bargmann structures
\begin{equation}
\rg=\delta_{ij}\,dx^i\otimes{}dx^j+dt\otimes\omega+\omega\otimes{}dt
\qquad
\&
\qquad
\xi=\frac{\partial}{\partial s}.
\label{BargmannTer}
\end{equation}
with $\omega$ as in (\ref{omega}), the Newton field equations (\ref{NewtonFieldEqs}) are of the form
\begin{equation}
\delta\bOmega=0
\qquad
\&
\qquad
\Delta_{\bR^\d}U+\frac{\partial}{\partial{}t}\delta\bomega+\half\Vert\bOmega\Vert^2=4\pi{}G\varrho
\label{NewtonCoriolis}
\end{equation}
where $\bomega=\omega_i(\bx,t)dx^i\vert_{\Sigma_t}$, and 
$\bOmega=d\bomega$ 
is the ``Coriolis'' $2$-form in the chosen (rotating) coordinates; we denote by $\delta$ the codifferential acting on differential forms of Euclidean space~$\Sigma_t\cong\bR^\d$. (In (\ref{NewtonCoriolis}), we have used the shorthand notation $\Vert\bOmega\Vert^2=\half\delta^{ik}\delta^{j\ell}\Omega_{ij}\Omega_{k\ell}$.) See also \cite{DGH} for the field equations in the general case given by Equations~(\ref{Bargmann}) and (\ref{gSigma}). 

\section{The Schr\"odinger-Newton equation}\label{SNSection}

\subsection{The Bargmann lift of the \SN equation}


The \SN (SN) field equations will be recast as follows on our Bargmann extension $(M,\rg,\xi)$ of NC spacetime. 

Consider first the Schr\"odinger equation. Let us declare ``wave functions'' to be complex-valued densities $\Psi$ of $M$ with weight 
\begin{equation}
w=\frac{\tn-2}{2\tn}.
\label{w}
\end{equation}
Recall that a $w$-density (with $w$ a complex number) is locally of the form $\Psi=\psi(x)\vert\Vol\vert^w$ where $\psi$ is a smooth complex-valued function and $\Vol$ some volume element of $M$; in our setting, we naturally choose the canonical volume element, $\Vol(\rg)$, on $(M,\rg)$, so that
\begin{equation}
\Psi=\psi(x)\vert\Vol(\rg)\vert^w.
\label{Psi}
\end{equation}

\bigskip 
$\bullet$ The \textit{Schr\"odinger equation} \cite{DBKP,DGH} then takes, along with the mass-constraint~(\ref{Mass}), the form of the wave equation
\begin{equation}
\medbox{
\Delta_Y(\rg)\Psi=0
}
\label{Yamabe}
\end{equation}
where $\Delta_Y(\rg)=\Delta(\rg)-(\tn-2)/(4(\tn-1))R(\rg)$ stands for the conformally-invariant Yamabe operator \cite{Bes,DO} sending $w$-densities to $(1-w)$-densities of $M$, with $w$ necessarily as in~(\ref{w}); see (\ref{QH}). Here, $\Delta(\rg)$ is the Laplace-Beltrami operator, and $R(\rg)$ the scalar curvature of $(M,\rg)$.
The wave functions, $\Psi$, are further assumed to be eigenvectors of the 
\textit{mass operator}, namely\footnote{The LHS of (\ref{Mass}) features the Lie derivative of a $w$-density, $\Psi$, of $M$ with respect to the vector field~$\xi$.}
\begin{equation}
\medbox{
\frac{\hbar}{i}L_\xi\Psi=m\Psi
}
\label{Mass}
\end{equation}
where $m>0$ stands for the mass of the quantum system; see Equation (\ref{Qm}). (This fixed eigenvalue will, later on, be licensed to undergo rigid dilations.)

\bigskip 
$\bullet$ We are ready to formulate the last ingredient of the (generalized) \textit{\SN equations} by deciding that $\varrho$ be in fact the quantum mass-density of the system, see Equation (\ref{varrho}). In view of the \textit{NC field equations} (\ref{NewtonFieldEqs}), we will therefore posit
\begin{equation}
\medbox{
\Ric(\rg)=4\pi{}G\,m\,\vert\tPsi\vert^2\,\theta\otimes\theta
}
\label{NewtonSch}
\end{equation}
where 
\begin{equation}
\tPsi=\frac{\Psi}{\Vert\Psi\Vert_\rg}
\qquad
\&
\qquad
\Vert\Psi\Vert^2_\rg= 
\int_{\Sigma_t}{\!\!\vert\psi\vert^2\,\vol(\hh)}<+\infty
\label{tPsi}
\end{equation}
is the corresponding normalized wave-function; here $\vol(\hh)$ stands for the canonical volume form of $\Sigma_t$,\footnote{This volume form can be defined intrinsically. Indeed, call $\eta=\rg^{-1}(\omega)$ the vector field associated with the connection form $\omega$ given by (\ref{omega}); one checks that $\eta$ is null and $\omega$-horizontal. Then $\Vol(\rg)(\xi,\eta)$ flows down to NC spacetime, $\cM$; once pulled-back to $\Sigma_t$, it canonically defines the volume $\d$-form $\vol(\hh)$. The latter admits the following local expression, namely $\vol(\hh)=\sqrt{\det(\rg_{ij}(x,t))}\,dx^1\wedge\cdots\wedge{}dx^\d$, where $\hh=\hh^{ij}(x,t)\,\partial/\partial{x^i}\otimes\partial/\partial{x^j}$ and $(\hh^{ij})=(\rg_{ij})^{-1}$.} and $\psi(\bx,t)=\psi_t(\bx)$ --- with $\psi_t\in{}L^2\left(\Sigma_t,\vol(h)\right)$ --- is as in (\ref{Psi}).

\goodbreak


\subsection{Recovering the SN equation in its standard guise}
\label{StandardSNsection}

We now claim that the coupled system of PDE (\ref{Yamabe}), (\ref{Mass}),(\ref{NewtonSch}) constitutes an intrinsic generalization of the Bargmann lift of the (non-local) \SN equation (\ref{SNEq}). Let us justify this statement in the special case where $M=\bR^\d\times\bR\times\bR$, the spatial metric being Euclidean, and the connection form (\ref{omega}) given by $\omega = -U(\bx,t)dt+ds$. In this case, the Bargmann structure (\ref{Bargmann}) reads
\begin{equation}
\rg=\delta_{ij}\,dx^i\otimes{}dx^j+dt\otimes{}ds+ds\otimes{}dt-2U(\bx,t)dt\otimes{}dt
\qquad
\&
\qquad
\xi=\frac{\partial}{\partial s}.
\label{BargmannBis}
\end{equation}
We find $\Ric(\rg)=(\Delta_{\bR^\d}U)dt\otimes{}dt$, and the Newtonian gravitational field equations (\ref{NewtonFieldEqs}) reduce to the ordinary Poisson equation (\ref{Poisson}); see also (\ref{NewtonCoriolis}) with $\bomega=0$.
Also does Equation~(\ref{Mass}) imply that 
\begin{equation}
\Psi(\bx,t,s)=e^{ims/\hbar}\,\psi(\bx,t)\vert\Vol(\rg)\vert^{\frac{\d}{2\d+4}}.
\label{Psi(bx,t,s)}
\end{equation}
 The wave-function~$\Psi$ being Yamabe-harmonic (see (\ref{Yamabe})), we get $\Delta_{\bR^\d}\Psi+2\partial/\partial{t}(\partial/\partial{s}\Psi)+2U(\partial/\partial{s})^2\Psi=0$, and hence $-\hbar^2/(2m)\Delta_{\bR^\d}\psi-i\hbar\,\partial\psi/\partial{t}+V\psi=0$, where $V=mU$ is the potential energy in the Schr\"odinger equation (\ref{SchEq}). At last, putting $V=\Delta_{\bR^\d}^{-1}(4\pi{}G\,m\varrho)$ with $\varrho$ as in (\ref{varrho}), enables us to recover the \SN equation (\ref{SNEq}) in its original form.

\subsection{The generalized \SN equation}
\label{GeneralizedSNEq}

Let us now take advantage of the geometrical form (\ref{Yamabe}), (\ref{Mass}), and (\ref{NewtonSch}) of the \SN equation to work out the brand new form of the SN equation in the spatially flat Bargmann structure (\ref{BargmannTer}), defined in terms of the Coriolis (co)vector potential $\bomega$ and the Newtonian potential~$U$. 

\goodbreak

Easy calculation yields the non-trivial components of the inverse metric, $\rg^{-1}$; they read as $\rg^{ij}=\delta^{ij}$, $\rg^{is}=-\omega_i$, $\rg^{ts}=1$, and $\rg^{ss}=2U+\delta^{ij}\omega_i\omega_j$, for all $i,j=1,\ldots,\d$. With the same sloppy notation, we express the non-zero Christoffel symbols as $\Gamma^i_{jt}=-\half\Omega_{ij}$, $\Gamma^i_{tt}=\partial_iU+\partial_t\omega_i$, $\Gamma^s_{ij}=\partial_{(i}\omega_{j)}$, $\Gamma^s_{it}=-\partial_iU-\half\Omega_{ij}\omega^j$, and $\Gamma^s_{tt}=-\partial_tU-\omega^i(\partial_iU+\partial_t\omega_i)$. 

With these data, the wave equation (\ref{Yamabe}) together with (\ref{Mass}) --- or (\ref{Psi(bx,t,s)}) --- leads to the Schr\"odinger equation 
\begin{equation}
\medbox{
-\frac{\hbar^2}{2m}\Delta_{\bR^\d}\psi
-\frac{\hbar}{2i}\left[\omega^j\circ\frac{\partial}{\partial{}x^j}+\frac{\partial}{\partial{}x^j}\circ\omega^j\right]\psi
+\frac{\hbar}{i}\frac{\partial\psi}{\partial{}t}
+m\left[U+\half\Vert\bomega\Vert^2\right]\psi
=0.
}
\label{SchEqBis}
\end{equation}
This equation and the coupled  NC field equations (\ref{NewtonSch}) \& (\ref{NewtonCoriolis}), namely
\begin{equation}
\medbox{
\delta\bOmega=0
\qquad
\&
\qquad
\Delta_{\bR^\d}U+\frac{\partial}{\partial{}t}\delta\bomega+\half\Vert\bOmega\Vert^2=4\pi{}G\,m\,\vert\tpsi\vert^2
}
\label{NewtonCoriolisBis}
\end{equation}
constitute the \textit{generalized \SN equations} resulting naturally from our choice of a Bargmann framework.


\section{Scaling symmetries of the SN equation}\label{SNSymmetries}

This section is devoted to the search of the symmetries of the SN equation under dilations using the intrinsic framework provided by Bargmann structures over NC spacetimes. 

\subsection{Conformally related Bargmann structures}

In view of Equation (\ref{Yamabe}), the symmetries of the previous coupled system will be naturally sought inside the conformal transformations of $(M,\rg)$, namely inside the pseudo-group of those local diffeomorphisms $\Phi$ of $M$ such that
\begin{equation}
\Phi^*\rg=\lambda_\Phi\,\rg
\label{Conf}
\end{equation}
with $\lambda_\Phi>0$ a smooth function of $M$.

\goodbreak

Moreover, we want that these transformations, $\Phi$, on our Bargmann manifold do actually project as \textit{bona fide} transformations of NC spacetime (on which the original SN equation was formulated). This implies that those $\Phi$ be mere automorphisms of the fibration $\pi:M\to{}\cM$, i.e., be such that
\begin{equation}
\Phi^*\xi=\nu_\Phi\,\xi
\label{AutFib}
\end{equation}
with $\nu_\Phi\neq0$ a smooth function of $M$. (Note that fibre-bundle automorphisms should require $\nu_\Phi=1$.)

One will show below (see also \cite{BDP3}) that, under these circumstances, $\lambda_\Phi$ is the pull-back of a (positive) smooth function of the time-axis, $T$, and $\nu_\Phi$ a (nonzero) constant, viz.,
\begin{equation}
d\lambda_\Phi\wedge\theta=0
\qquad
\&
\qquad
d\nu_\Phi=0.
\label{lambdanu}
\end{equation}
Indeed, if $\hrg=\Phi^*\rg$ and $\hxi=\Phi^*\xi$ are as in (\ref{Conf}) and (\ref{AutFib}) respectively, we demand that $(M,\hrg,\hxi)$ be again Bargmann. The vector field $\hxi$ being automatically $\hrg$-null, it remains to find under which conditions it is parallel transported by the Levi-Civita connection $\hnabla$ of~$\hrg$. We find that $\hnabla\hxi=0$ iff $d\nu_\Phi\otimes\theta+\nu_\Phi/(2\lambda_\Phi)\left[\xi(\lambda_\Phi)\,\rg+d\lambda_\Phi\wedge\theta\right]=0$. Straightforward calculation shows that $d\nu_\Phi=(\nu_\Phi/\lambda_\Phi)d\lambda_\Phi+f\theta$ for some function $f$, as well as $\xi(\lambda_\Phi)=0$. This, in turn, implies $d\lambda_\Phi\wedge\theta=0$, hence $d\nu_\Phi=0$. We have just proved (\ref{lambdanu}).

From now on, and to avoid clutter, we will write $\lambda$ (resp. $\nu$) 
instead of $\lambda_\Phi$ (resp. $\nu_\Phi$) if no confusion can occur.

\subsection{A Schwarzian intermezzo}\label{Intermezzo}

Consider, on a pseudo-Riemannian manifold $(M,\rg)$, a metric $\widehat\rg=\lambda\,\rg$  conformally related to  $\rg$. We will denote by $\Ric$ the mapping from the set of pseudo-Riemannian metrics of $M$ to the space of twice covariant symmetric tensor fields of $M$ defined by the Ricci tensor. 

\goodbreak

The conformal transformation law of the Ricci tensor is well-known and reads
\begin{equation}
\Ric(\hrg)
=
\Ric(\rg)
-\frac{(\tn-2)}{2}\left(\frac{\nabla{d\lambda}}{\lambda}
-
\frac{3}{2}\,\frac{d\lambda\otimes{}d\lambda}{\lambda^2}\right)
-
\half\left(\frac{\Delta{\lambda}}{\lambda}
+
\frac{(\tn-4)}{2}\,\frac{\vert{}d\lambda\vert^2}{\lambda^2}\right)\rg
\label{RicciHat}
\end{equation}
where $\Delta{\lambda}=\rg^{\alpha\beta}\nabla_\alpha\partial_\beta\lambda$ and
$\vert{}d\lambda\vert^2=\rg^{\alpha\beta}\partial_\alpha\lambda\partial_\beta\lambda$, in a coordinate system $(x^\alpha)_{\alpha=1,\ldots,\tn}$ of $M$. The scalar curvature then transforms as
\begin{equation}
R(\hrg)
=
\frac{R(\rg)}{\lambda} -(\tn-1)\left(
\frac{\Delta{\lambda}}{\lambda^2}+\frac{(\tn-6)}{4}\,\frac{\vert{}d\lambda\vert^2}{\lambda^3}
\right).
\label{RHat}
\end{equation}

\goodbreak

Specialize then considerations to Bargmann structures $(M,\rg,\xi)$.
It has just been proved, see also \cite{DL}, that if $(M,\hrg,\hxi)$ where $\hrg=\lambda\rg$ and $\hxi=\nu\xi$ (as in (\ref{Conf}) and (\ref{AutFib}))
is again Bargmann structure, then
$\lambda:t\mapsto{}\lambda(t)$ is (the pull-back of) a function of the time-axis $T$. We then have $d\lambda=\lambda'\theta$, and hence
$R(\hrg)=R(\rg)/\lambda$
since $\Delta \lambda=\lambda''\rg^{\alpha\beta}\theta_\alpha\theta_\beta=0$ and $\vert{}d\lambda\vert^2=(\lambda')^2\rg^{\alpha\beta}\theta_\alpha\theta_\beta=0$; cf. Equation (\ref{RHat}).

\goodbreak

Introduce now $\varphi\in\Diff_+(T)$ via
$\lambda(t)=\varphi'(t)$;
then Equation (\ref{RicciHat}) can be cast into the quite remarkable form, viz.,
$\Ric(\hrg)
=
\Ric(\rg)
-\half(\tn-2)\,S(\varphi)\,\theta\otimes\theta$,
where
\begin{equation}
S(\varphi)
=
\frac{\varphi'''}{\varphi'}
-
\frac{3}{2}\left(\frac{\varphi''}{\varphi'}\right)^2
\label{Schwarzian}
\end{equation}
is the \textit{Schwarzian derivative} of $\varphi$. Notice that the latter naturally shows up as a \textit{quadratic differential} of $T$, namely $\cS(\varphi)=S(\varphi)\,\theta\otimes\theta$ where
\begin{equation}
\cS(\varphi)=\frac{-2}{\tn-2}\left(\Ric(\hrg)-\Ric(\rg)\right).
\label{SchwarzianBis}
\end{equation}

\subsection{Rescalings in action}

Apart from the natural actions of the group of diffeomorphisms on our geometric objects (see Section \ref{NaturalSection}), we will first need the transformation of the latter under rescaling. Those will ultimately enter the definition of the group of Bargmann conformal transformations, leading to the group of SN symmetries under study.

\goodbreak

\bigskip 
$\bullet$ It is a classical result (see, e.g., \cite{MRS}) that under a conformal rescaling 
\begin{equation}
\rg\mapsto\hrg=\lambda\rg
\qquad
\&
\qquad
\lambda\in{}C^\infty(M,\bR^*_+)
\label{hrg}
\end{equation}
of the metric, the Yamabe operator enjoys the following invariance property, viz.,
\begin{equation}
\medbox{
\Delta_Y(\hrg)=\Delta_Y(\rg)
}
\label{Yamabe=ConfInv}
\end{equation}
being clearly understood that the Yamabe operator, $\Delta_Y(\rg)$, maps here $w$-densities to $(1-w)$-densities of $M$, where the weight $w$ is as in (\ref{w}).\footnote{
Let us recall that we have
\begin{equation}
\Delta_Y(\hrg)
=
\lambda^{-\frac{\tn+2}{4}}\circ\Delta_Y(\rg)\circ\lambda^{\frac{\tn-2}{4}}
\label{hatYamabe}
\end{equation}
if the Yamabe operator is rather viewed as an operator acting on $0$-densities, i.e., smooth functions of $M$.
}
\goodbreak



\bigskip 
$\bullet$ We will now prove that the normalized wave function $\tPsi$ transforms as
\begin{equation}
\htPsi=
\lambda^{-\frac{\d}{4}}\tPsi
\label{htPsi}
\end{equation}
and that the latter holds true for any $w$-density $\Psi$. Indeed, in view of (\ref{tPsi}), we find that $\htPsi=\lambda^\frac{Nw}{2}\Psi/\Vert\lambda^\frac{Nw}{2}\Psi\Vert_\hrg=\Psi/\Vert\Psi\Vert_\hrg$ since $\lambda$ is a function of $T$. Now, in Equation (\ref{tPsi}) we have $\vol(\hath)=\lambda^\frac{\d}{2}\vol(\hh)$ so that $\Vert\Psi\Vert_\hrg=\lambda^\frac{\d}{4}\Vert\Psi\Vert_\rg$; this ends the proof of Equation (\ref{htPsi}).

\bigskip 
$\bullet$ Under a rescaling $\xi\mapsto\hxi=\nu\xi$, we clearly have
\begin{equation}
L_\hxi = \nu L_\xi.
\end{equation}

\bigskip 
$\bullet$  Let us recall that, in Section \ref{Intermezzo}, we have already shown that
\begin{equation}
\medbox{
\Ric(\hrg)
=
\Ric(\rg)
-\frac{(\tn-2)}{2}\,S(\varphi)\,\theta\otimes\theta
}
\label{RicciHatBis}
\end{equation}
holds true whenever $(M,\rg,\xi)$ and $(M,\hrg,\hxi)$ are Bargmann manifolds (we remember that $\lambda(t)=\varphi'(t)>0$ for all $t\in{}T$).

\subsection{The dilation-invariance of the SN equation}

Let us first remind that, for the wave equation (\ref{Yamabe}), pure conformal rescalings (\ref{hrg}) of the metric translate as follows: if $(\rg,\Psi)$ is a solution of the wave equation $\Delta_Y(\rg)\Psi=0$, then Equation (\ref{Yamabe=ConfInv}) implies that the same is true for $(\hrg,\Psi)$, namely that $\Delta_Y(\hrg)\Psi=0$.

\medskip

Likewise, in the case under study, we wish to determine under which conditions $(\hrg,\hxi,\Psi)$ is a solution of the SN equations (\ref{Yamabe})--(\ref{NewtonSch}) if $(\rg,\xi,\Psi)$ is such a solution.


\bigskip 
$\bullet$ Using the above argument about the fundamental property (\ref{Yamabe=ConfInv}) of the Yamabe operator, we readily get
\begin{equation}
\Delta_Y(\rg)\Psi=0
\qquad
\implies
\qquad
\Delta_Y(\hrg)\Psi=0
\label{YamabeEqConfInv}
\end{equation}
confirming that the wave equation (\ref{Yamabe}) is indeed conformally invariant.


\bigskip 
$\bullet$ As to the behavior of the mass under a dilation $\xi\mapsto\hxi=\nu\xi$, we claim that 
\begin{equation}
\frac{\hbar}{i}L_\xi\,\Psi=m\Psi
\qquad
\implies
\qquad
\frac{\hbar}{i}L_\hxi\,\Psi=\hm\Psi
\label{mDilation}
\end{equation}
and thus prove that the mass (as defined by (\ref{Mass})) gets dilated according to \cite{BGDB,GG}
\begin{equation}
\medbox{
\hm=\nu\,m
}
\label{hm}
\end{equation}
in coherence with Equation (\ref{hmTer}) in Appendix \ref{MassDilation}.

\goodbreak


\bigskip 
$\bullet$ Let us write the Newton-Cartan equation (\ref{NewtonSch}) in terms of the dilated objects. Using Equations~(\ref{hm}), (\ref{htPsi}), and the fact that dilations act on the clock as
\begin{equation}
\htheta=\lambda\nu\,\theta
\label{hclock}
\end{equation}
we readily find that
\begin{eqnarray}
\nonumber
0&=&\Ric(\hrg)-4\pi G\hm\,\big\vert\htPsi\big\vert^2\,\htheta\otimes\htheta\\[4pt]
\nonumber
&=&\Ric(\hrg)-4\pi G m\,\nu^3\lambda^{2-\frac{\d}{2}}\big\vert\tPsi\big\vert^2\,\theta\otimes\theta\\
&=&\Ric(\rg)-4\pi G m\,\big\vert\tPsi\big\vert^2\,\theta\otimes\theta\\
&&-\left[\frac{\d}{2}\,S(\varphi)+4\pi G m\,(\nu^3\lambda^{2-\frac{\d}{2}}-1)\big\vert\tPsi\big\vert^2\right]\theta\otimes\theta
\end{eqnarray}
in view of (\ref{RicciHatBis}). 

\goodbreak

We can thus immediately conclude that
\begin{equation}
\Ric(\rg)=4\pi G m\,\big\vert\tPsi\big\vert^2\theta\otimes\theta\qquad
\implies
\qquad
\Ric(\hrg)=4\pi G\hm\,\big\vert\htPsi\big\vert^2\,\htheta\otimes\htheta\label{hatNewton}
\end{equation}
provided, on the one hand,\footnote{We thus get that $\varphi\in\PGL(2,\bR)$.}
\begin{equation}
\medbox{
S(\varphi)=0
}
\label{S=0}
\end{equation}
since $\varphi$ is an orientation-preserving diffeomorphism of the time-axis, $T$, while~$\tPsi$ is defined on the whole Bargmann manifold, $M$, and on the other hand,
\begin{equation}
\medbox{
\lambda^{2-\frac{\d}{2}}\,\nu^3=1.
}
\label{dilationConstraint}
\end{equation} 

\bigskip 
$\bullet$ The fundamental constraint (\ref{dilationConstraint}) thus implies that\begin{equation}
\lambda=\nu^{-\frac{6}{4-\d}}
\label{lambda}
\end{equation}
is a (strictly positive) constant,\footnote{We discover that, thanks to (\ref{hm}), these dilations preserve the \textit{positivity of the mass} since $\nu\in\bR^*_+$.} and that $\varphi$ is actually an orientation-preserving \textit{affine diffeomorphism} of $T$, namely
\begin{equation}
\varphi(t)=\lambda t+\mu
\qquad
\&
\qquad
\d\neq4
\label{varphi}
\end{equation}
with $\lambda\in\bR^*_+$, and $\mu\in\bR$. Inversions are lost in this case

\bigskip 
$\bullet$ If $\d=4$, we see that (\ref{dilationConstraint}) yields $\nu=1$ (trivial mass-dilation (\ref{hm})); this proves that, in this case, we have
\begin{equation}
\varphi(t)=\frac{d t+e}{f t+g}
\qquad
\&
\qquad
dg-ef=1
\label{varphid=4}
\end{equation}
or, equivalently, $\varphi\in\PSL(2,\bR)$.

\section{Maximal group of invariance of the SN equation}\label{SNgroupSection}

\subsection{Naturality relationships}\label{NaturalSection}

Let us recall that, on any (pseudo-)Riemannian manifold $(M,\rg)$ we have the \textit{naturality} relationship of the Ricci mapping, viz.,
\begin{equation}
\Ric(\Phi^*\rg)=\Phi^*(\Ric(\rg))
\label{PhiRic=RicPhi}
\end{equation}
for all $\Phi\in\Diff(M)$; see Equation (5.13) in \cite{Bes}.

\goodbreak

Likewise, we know that
\begin{equation}
\Delta_Y(\Phi^*\rg)=\Phi^*(\Delta_Y(\rg))
\label{PhiYamabe=YamabePhi}
\end{equation}
for all $\Phi\in\Diff(M)$; see, e.g, \cite{MRS} and references therein.

At last, the following holds true
\begin{equation}
L_{\Phi^*\xi}=\Phi^*(L_\xi)
\label{PhiLie=LiePhi}
\end{equation}
for all $\Phi\in\Diff(M)$.

\subsection{\SN symmetries}

We are now ready to determine in geometric terms the symmetries of the \SN system. In contradistinction to the full Schr\"odinger symmetry of another system of coupled PDE, namely the Schr\"odinger-Chern Simons system in flat ($2+1$)-dimensional Galilei spacetime \cite{JP,DHP1}, the symmetry of the present system turns out to be more subtle, and intricate to work out, since the system is in fact gravitationally ``self-coupled''.

\goodbreak

As usual, we call $\Conf(M,\rg)$ the ``group'' of (signature-preserving) conformal transformations of the metric $\rg$ of $M$, the \textit{maximal invariance group} of the wave equation~(\ref{Yamabe}). 

Let us recall that the chronoprojective ``group'' \cite{Duv1,Duv2,BDP1,BDP2} of a Bargmann structure is generated by conformal diffeomorphisms of $(M,\rg)$ that are also automorphisms of the fibration $\pi:M\to{}\cM=M/(\bR\xi)$, i.e.,
\begin{eqnarray}
\Chr(M,\rg,\xi)
&=&\Conf(M,\rg)\cap\Aut(M\to{}\cM)\\
&=&
\{\Phi\in\Diff(M)\strut\,\vert\,\Phi^*\rg=\lambda\rg,\Phi^*\xi=\nu\xi\}
\label{Chr}
\end{eqnarray}
where necessarily (see (\ref{lambdanu})) $\lambda\in{}C^\infty(T,\bR^*_+)$, and $\nu\in\bR^*$.

\goodbreak

Our aim is to determine in full generality the subgroup, $\rSN(M,\rg,\xi)$, of  $\Chr(M,\rg,\xi)$ 
that permutes the solutions, $(\rg,\xi,\Psi)$, of the \SN equations (\ref{Yamabe})--(\ref{NewtonSch}), i.e., that maps solutions to solutions of this system.


\bigskip 
$\bullet$ Assuming $\Psi$ satisfies $\Delta_Y(\rg)\Psi=0$, we find that Equations (\ref{PhiYamabe=YamabePhi}) 
entails
\begin{eqnarray}
\nonumber
0&=&\Phi^*(\Delta_Y(\rg)\Psi)\\
&=&\Delta_Y(\Phi^*\rg)\Phi^*\Psi
\end{eqnarray}
which clearly confirms the $\Diff(M)$-naturality of the wave equation (\ref{Yamabe}). 

\goodbreak

Moreover its turns out that (\ref{Yamabe=ConfInv}) implies
\begin{equation}
\Delta_Y(\rg)\Phi^*\Psi=0
\end{equation}
hence that if $\Psi$ is Yamabe-harmonic, so is $\Phi^*\Psi$ for all $\Phi\in\Conf(M,\rg)$.


\bigskip 
$\bullet$ Likewise, using Equations (\ref{PhiLie=LiePhi}) and (\ref{PhiStarm}), we find that
\begin{eqnarray}
\nonumber
0&=&\Phi^*\left(-i\hbar\,L_\xi\Psi-m\,\Psi\right)\\
&=&(-i\hbar\,L_{\Phi^*\xi})\Phi^*\Psi-\hm\,\Phi^*\Psi
\label{NaturalityOfMass}
\end{eqnarray}
which guarantees the $\Diff(M)$-naturality of the mass equation (\ref{Mass}). 

\goodbreak

Together with (\ref{AutFib}) and (\ref{hm}), Equation (\ref{NaturalityOfMass}) leads to
\begin{equation}
\frac{\hbar}{i}\,L_\xi(\Phi^*\Psi)=m\,\Phi^*\Psi
\label{mPhiStarPsi=m}
\end{equation}
proving that $\Phi^*\Psi$ has also mass $m$ for all $\Phi\in\Chr(M,\rg,\xi)$.

\goodbreak


\bigskip 
$\bullet$ We now turn to the classical NC field equation coupled to the Schr\"odinger equation, namely Equations (\ref{Yamabe}) and (\ref{Mass}). 

Let $(\rg,\xi,\Psi)$ be one of the solutions of (\ref{NewtonSch}). With the help of (\ref{PhiRic=RicPhi}) we find that 
\begin{eqnarray}
0&=&
\nonumber
\Phi^*\big(\Ric(\rg)-4\pi{}G\,m\,\vert\tPsi\vert^2\,\theta\otimes\theta\big)\\
&=&
\nonumber
\Phi^*\big(\Ric(\rg)\big)-4\pi{}G\,\hm\,\vert\Phi^*\tPsi\vert^2\,\Phi^*\theta\otimes\Phi^*\theta\\
&=&
\Ric(\Phi^*\rg)-4\pi{}G\,\hm\,\vert\Phi^*\tPsi\vert^2\,\Phi^*\theta\otimes\Phi^*\theta
\label{NaturalityNC}
\end{eqnarray}
which expresses the $\Diff(M)$-naturality of the NC field equations (\ref{NewtonSch}).

Besides, it is an easy matter to show that
\begin{equation}
\Phi^*\tPsi
=
\lambda^{-\frac{\d}{4}}\,\widetilde{\Phi^*\Psi}
\label{PhiStartPsi}
\end{equation}
for all $\Phi\in\Chr(M,\rg,\xi)$. Indeed, from (\ref{Psi}), and putting $(\hx,\wht,\hs)=\Phi(x,t,s)$, we get\footnote{We use systematically the fact that, in the chosen local coordinate system $M$, we have 
\begin{equation}
\sqrt{\det(\rg_{ij}(\hx,\wht))}\det\left(\frac{\partial\hx^i}{\partial{}x^j}\right)=\lambda^{\frac{\d}{2}}\sqrt{\det(\rg_{ij}(x,t))}.
\label{Jacobian}
\end{equation}
}
\begin{eqnarray}
\nonumber
\Vert\Phi^*\Psi\Vert^2_\rg
&=&
\int_{\Sigma_t}{
\!\vert\lambda^{\frac{\d}{4}}\,\Phi^*\psi\vert^2\,\vol(h)
}\\
&=&
\nonumber
\int_{\Sigma_t}{
\!\vert\psi(\hx,\wht,\hs)\vert^2\,\lambda^{\frac{\d}{2}}\sqrt{\det(\rg_{ij}(x,t))}\,dx^1\ldots{}dx^\d
}\\
&=&
\nonumber
\int_{\Sigma_\wht}{
\!\vert\psi(\hx,\wht,\hs)\vert^2\sqrt{\det(\rg_{ij}(\hx,\wht))}\,d\hx^1\ldots{}d\hx^\d
}\\
&=&
\Vert\Psi\Vert^2_\rg.
\label{Unitarity}
\end{eqnarray}
According to (\ref{tPsi}), we find
$\Phi^*\tPsi
=
\Phi^*\Psi/\Vert\Phi^*\Psi\Vert_{\Phi^*\rg}
=
\Phi^*\Psi/(\lambda^{\frac{\d}{4}}\Vert\Psi\Vert_\rg)$, 
by using again~(\ref{Jacobian}). With the help of (\ref{Unitarity}), we then write $\Phi^*\tPsi
=\lambda^{-\frac{\d}{4}}\,\Phi^*\Psi/\Vert\Phi^*\Psi\Vert_\rg
$, proving Equation~(\ref{PhiStartPsi}).

Equations (\ref{NaturalityNC}) and (\ref{PhiStartPsi}), together with (\ref{RicciHatBis}) and (\ref{S=0}) then yield in a straightforward manner
\begin{equation}
\Ric(\rg)=4\pi{}G\,m\,\vert\widetilde{\Phi^*\Psi}\vert^2\,\theta\otimes\theta
\label{NCBis}
\end{equation}
where (\ref{dilationConstraint}) has been duly taken into account. 

\goodbreak

We stress that $\Phi^*\Psi$ thus also provides a solution of the NC gravitational field equation (\ref{NewtonSch}), for any \SN diffeo\-morphism~$\Psi\in\rSN(M,\rg,\xi)$ of the Bargmann manifold, i.e., any $\Phi\in\Chr(M,\rg,\xi)$ for which~(\ref{dilationConstraint}) holds true.

\subsection{The general definition of the \SN group}

Taking advantage of both the definition (\ref{Chr}) of the chronoprojective group, and the constraint (\ref{dilationConstraint}), we can now define in quite general terms the \SN group of a Bargmann structure as
\begin{equation}
\medbox{
\rSN(M,\rg,\xi)=
\{\Phi\in\Diff(M)\strut\,\vert\,\Phi^*\rg=\lambda\rg,\Phi^*\xi=\nu\xi,\,\lambda^{2-\frac{\d}{2}}\,\nu^3=1\}.
}
\label{SN}
\end{equation}
We notice that this definition makes sense if $\lambda>0$, hence~$\nu>0$.

If $\d=4$, Equation (\ref{SN}) yields $\nu=1$; the \SN group is, in this case, isomorphic to the Schr\"odinger group (\ref{Sch}), namely
\begin{equation}
\rSN(M,\rg,\xi)=\Sch(M,\rg,\xi)
\qquad
\hbox{if}
\qquad
\d=4.
\label{SN=Sch}
\end{equation}

If $\d\neq4$ the full Schr\"odinger symmetry group (\ref{Sch}) is thus broken to the \SN group (\ref{SN}).\footnote{The strange, singular, dimension $\d=4$ plays a r\^ole akin to that, $\tn$, of a relativistic spacetime~$(M,\rg)$ for which the Maxwell Lagrangian density $L(F,\rg)=\frac{1}{4}\rg^{\alpha\beta}\rg^{\gamma\delta}F_{\alpha\gamma}F_{\beta\delta}\,\vert\Vol(\rg)\vert$ is conformally invariant. Indeed, $L(F,\lambda\rg)=\lambda^{-(2-\frac{\tn}{2})}L(F,\rg)$ implies $L(F,\lambda\rg)=L(F,\rg)$ for all $\lambda\in{}C^\infty(M,\bR^*_+)$ if $\tn=4$. Here, it is the conformal symmetry which is broken whenever $\tn\neq4$.\label{footnoteMaxwell}}

\goodbreak

\subsection{The \SN group for the spatially-flat model}

Let us now determine explicitly the SN group (\ref{SN}) of invariance of the \SN equation  (\ref{SNEq}). We will choose, however, to deal with the most general spatially-flat Bargmann metric (\ref{BargmannTer}) and the associated generalized \SN equation given by Equations (\ref{Yamabe}), (\ref{Mass}) and (\ref{NewtonSch}); as shown in Section \ref{StandardSNsection}, the case of the standard \SN equation will merely follow as a special case of the generalized SN equation given by (\ref{SchEqBis}) and (\ref{NewtonCoriolisBis}).

\goodbreak

\subsubsection{The chronoprojective group of the flat Bargmann structure}
\label{AppendixC}

The general solution of Equations (\ref{Conf}) and (\ref{AutFib}) for the flat Bargmann structure (\ref{FlatBargmann}) is given by the so-called ``chronoprojective group''\footnote{This group is in fact the canonical $(\bR,+)$-extension of the projected ``chronoprojective group'' acting on NC spacetime via (\ref{hbx}) and (\ref{wht}). One often uses the same term for both groups. Formula~(\ref{hs}) is a generalization to the case~$\nu\neq1$ of the corresponding one, for $\nu=1$, in \cite{DHP4}.} denoted $\Chr(\bR^{\d+1,1})=\Chr(\bR^{\d+2},\rg_0,\xi)$ \cite{Duv1,Duv2,BDP1,BDP2,BDP3}. 

Its projective action $\Phi:(\bx,t,s)\mapsto(\hbx,\wht,\hs)$ on $\bR^{\d+2}$ reads\footnote{Below, $\la\,\cdot\,,\,\cdot\,\ra$ stands for the Euclidean scalar product of $\bR^\d$, and $\Vert\,\cdot\,\Vert$ for the associated norm.}
\begin{eqnarray}
\label{hbx}
\hbx&=&\frac{A\bx+\bb t+\bc}{f t+g}\\[4pt]
\label{wht}
\wht&=&\frac{d t+e}{f t+g}\\[4pt]
\label{hs}
\hs&=&\frac{1}{\nu}\left[
s+\frac{f}{2}\frac{\Vert{}A\bx+\bb t+\bc\Vert^2}{f t+g}
-\la\bb,A\bx\ra-\half\Vert\bb\Vert^2t+h
\right]
\end{eqnarray}
where $A\in\rO(\d)$, $(\bb,\bc)\in\bR^\d\times\bR^\d$ (boosts \& space-translations), also
\begin{equation}
D=\left(
\begin{array}{cc}
d&e\\f&g
\end{array}
\right)\in\GL(2,\bR)
\label{GL2}
\end{equation}
represents the projective group of the time-axis (whence the name \textit{chronoprojective} transformation for (\ref{hbx})--(\ref{hs})), and $h\in\bR$ (extension parameter). 

We note that the factors
\begin{equation}
\lambda=\frac{1}{(ft+g)^2}
\qquad
\&
\qquad
\nu=dg-ef
\label{lambdanuBis}
\end{equation}
are duly expressed in terms of the group parameters. In particular, the coefficient $\nu$ in~(\ref{hs}) is given by (\ref{lambdanuBis}).
\goodbreak

\subsubsection{The SN group as a subgroup of the chronoprojective group}

Let us first determine the chronoprojective group $\Chr(\bR^{\d+2},\rg,\xi)$, where $\rg$ and $\xi$ are as in (\ref{BargmannTer}), namely
\begin{equation}
\rg=\rg_0+dt\otimes\omega_i(\bx,t)\,dx^i+\omega_i(\bx,t)\,dx^i\otimes{}dt-2U(\bx,t)\,dt\otimes{}dt
\qquad
\&
\qquad
\xi=\frac{\partial}{\partial s}
\label{BargmannQuater}
\end{equation}
where $\rg_0$
is the flat Bargmann metric (\ref{FlatBargmann}).

\goodbreak

We start by characterizing all $\Phi\in\Conf(\bR^{\d+2},\rg)$, i.e., such that 
\begin{eqnarray}
\nonumber
\Phi^*\rg
&=&
\Phi^*\rg_0+d\wht\otimes\homega_i\,d\hx^i+\homega_i\,d\hx^i\otimes{}d\wht
-2\hU\,d\wht\otimes{}d\wht\\
&=&
\lambda\rg_0+\lambda\left(dt\otimes\omega_i\,dx^i+\omega_i\,dx^i\otimes{}dt-2U\,dt\otimes{}dt\right)
\label{PhiStarg}
\end{eqnarray}
for some $\lambda\in{}C^\infty(\bR^{\d+2},\bR^*_+)$. Now we observe that the metric $\rg_0$ contains no terms of the form $dt\otimes{}dx^i$ (for $i=1,\ldots,\d$) or even $dt\otimes{}dt$. From this, we conclude that Equation~(\ref{PhiStarg}) insures that $\Phi^*\rg_0=\lambda\rg_0$, i.e., that $\Phi\in\Conf(\bR^{\d+2},\rg_0)$. 

We furthermore want that these diffeo\-morphisms, $\Phi$, satisfy Equation~(\ref{AutFib}), namely $\Phi^*\xi=\nu\xi$; this entails that the sought SN group is in fact a (proper) subgroup of the chronoprojective group of the flat Bargmann structure, 
\begin{equation}
\rSN(\bR^{\d+2},\rg,\xi)\subset\Chr(\bR^{\d+2},\rg_0,\xi).
\label{SNsubsetChr}
\end{equation}
These (local) diffeomorphisms $\Phi:(\bx,t,s)\mapsto(\hbx,\wht,\hs)$ are hence explicitly given by Equations~(\ref{hbx}), (\ref{wht}) and (\ref{hs}).

\goodbreak

\bigskip 
$\bullet$ If $\d\neq4$, we have proved in (\ref{varphi}), and in quite general terms, that $\wht=\lambda\,t+\mu$, where $\lambda\in\bR^*_+$ and $\mu\in\bR$. This entails that
\begin{equation}
f=0
\label{f=0}
\end{equation}  
in (\ref{lambdanuBis}), so that 
\begin{equation}
\lambda=\frac{1}{g^2}
\qquad
\&
\qquad
\nu=d\,g.
\label{lambdanuTer}
\end{equation}

Returning to the fundamental constraint (\ref{dilationConstraint}), we find that 
\begin{equation}
d=\nu^{\frac{\d-1}{\d-4}}
\qquad
\&
\qquad
g=\nu^{-\frac{3}{\d-4}}.
\label{dg}
\end{equation}
We thus claim that the \SN group is, in this case, isomorphic to the multiplicative group of those matrices of the form
\begin{equation}
\left(
\begin{array}{cccc}
A&\bb&0&\bc\\[4pt]
0&d&0&e\\[4pt]
\displaystyle
-\frac{\bb^tA}{d}&
\displaystyle
-\frac{\Vert\bb\Vert^2}{2d}&
\displaystyle
\frac{1}{d}&
\displaystyle
\frac{h}{d}\\[8pt]
0&0&0&g
\end{array}
\right)\in\rSN(\bR^{\d+2},\rg,\xi)
\label{SNinGL}
\end{equation}
where 
$A\in\rO(\d)$, $(\bb,\bc)\in\bR^\d\times\bR^\d$, $(e,h)\in\bR^2$, and $\nu\in\bR^*_+$, with $d$ and $g$ as in (\ref{dg}).

\goodbreak

Considering the only subgroup of dilations, we find that 
\begin{eqnarray}
\label{alpha}
\hbx&=&\nu^{\frac{3}{\d-4}}\bx\\
\label{beta}
\wht&=&\nu^{\frac{\d+2}{\d-4}}{}t\\
\label{gamma}
\hs&=&\nu^{-1}s
\label{dilations}
\end{eqnarray}
with $\nu\in\bR^*_+$, in full agreement with the claim in Reference \cite{GG} in the case $\d=3$.

We can, at this stage, compute the \textit{dynamical exponent}, $z$, of the SN group. Recall that it is defined by $\wht=\alpha^zt$ if $\hbx=\alpha\bx$ for a dilation $\alpha\in\bR^*_+\subset\GL(2,\bR)$; it measures how much is time dilated as compared to space.\footnote{The notion of dynamical exponent is specific to non-relativistic theories; see, e.g., \cite{Hen}. It has clearly no relativistic analogue.} Using (\ref{alpha}) and~(\ref{beta}), we readily find
\begin{equation}
\medbox{
z=\frac{\d+2}{3}.
}
\label{z}
\end{equation}

\goodbreak

\bigskip 
$\bullet$ If $\d=4$, we have already shown that the \SN group is in fact defined by the constraint $\nu=1$, i.e.,
\begin{equation}
dg-ef=1
\label{nu=1}
\end{equation}
in Equations (\ref{GL2}) and (\ref{lambdanuBis}). This confirms that, in this case, the \SN group is iso\-morphic to the Schr\"odinger group, i.e.,
\begin{equation}
\rSN(\bR^{5,1})\cong\Sch(\bR^{5,1}).
\label{SN51}
\end{equation}

We know that the dynamical exponent of the Schr\"odinger group, $\Sch(\bR^{\d+1,1})$ is $z=2$, in perfect accordance with our general expression (\ref{z}) in the case $\d=4$.

\subsubsection{The SN group as group of invariance of the NC equation}

Although we have already provided an intrinsic proof that Condition (\ref{dilationConstraint}) is the very constraint defining the SN group, we will nonetheless, and as an alternative, return to this proof using the specific NC field equations (\ref{NewtonCoriolis}) for the chosen spatially-flat Bargmann metric. 

To this end, let us compute explicitly the expression (\ref{PhiStarg}) of the transformed metric, $\Phi^*\rg$, using the general form of the chronoprojective transformations $(\bx,t)\mapsto(\hbx,\wht)$ given by (\ref{hbx}) and (\ref{wht}). In doing so, for any spatial dimension $\d$, we get the new Coriolis form, $\hbomega$, and Newtonian potential, $\hU$; owing to Equations  (\ref{f=0}) and (\ref{lambdanuTer}), we find
\begin{equation}
\hbomega=\lambda^{-\half}\nu^{-1}\bomega.A^{-1}
\qquad
\&
\qquad
\hU=\lambda^{-1}\nu^{-2}\left(U+\bomega.A^{-1}\bb\right).
\label{hbomegahU}
\end{equation}
Moreover, the expression (\ref{hbx}) of $\hbx$, yields\footnote{To alleviate the notation, we will put $\hpartial_i=\partial/\partial{}\hx^i$. We will also write $\grad F=\delta^{ij}\partial_iF\,\partial_j$.}
\begin{equation}
\hpartial_iF=\lambda^{-\half}(A^{-1})^j_i\,\partial_jF
\qquad
\&
\qquad
\hpartial_tF=\lambda^{-1}\nu^{-1}\left(\partial_tF-\grad{F}\cdot{}A^{-1}\bb\right)
\label{hpartial}
\end{equation}
for all $i=1,\ldots,\d$, and all function $F\in{}C^\infty(\bR^{\d+1},\bR)$ defined on spacetime. 
\goodbreak

Remembering that the codifferential (the divergence operator) of the $1$-form $\bomega$ is $\delta\bomega=\delta^{ij}\partial_i\omega_j$, we easily find (via~(\ref{hbomegahU}) and (\ref{hpartial})) that
\begin{equation}
\hdelta\,\hbomega=\lambda^{-1}\nu^{-1}\delta\bomega.
\label{hdeltahbomega}
\end{equation}
Easy calculation, using the fact that $\Delta_{\bR^\d}U=\delta^{ij}\partial_i\partial_jU$, the field equation $\delta\bOmega=0$, and Equation (\ref{hbomegahU}) also yields
\begin{equation}
\hDelta_{\bR^\d}\hU=\lambda^{-2}\nu^{-2}\left[\Delta_{\bR^\d}U+\grad(\delta\bomega).A^{-1}\bb\right]
\label{hDeltahU}
\end{equation}
where $\bb\in\bR^\d$ is a Galilean boost.

\goodbreak

Let us furthermore mention that $\Omega_{ij}=\lambda^{-1}\nu^{-1}(A^{-1})_i^k(A^{-1})_j^\ell\,\Omega_{k\ell}$ for all $i,j=1,\ldots,d$, which leads to
\begin{equation}
\Vert\hbOmega\Vert^2=\lambda^{-2}\nu^{-2}\Vert\bOmega\Vert^2.
\label{hbOmega2}
\end{equation}

It is now possible to determine under which condition do the NC (alias Newton-Coriolis) equations~(\ref{NewtonCoriolis}) remain invariant under a chronoprojective transformation.

\bigskip 
$\bullet$ Firstly, in view of the previous preparation, we clearly have
\begin{equation}
\hdelta\,\hbOmega=\delta\,\bOmega=0.
\label{hNC1}
\end{equation}

\bigskip 
$\bullet$ Secondly, making use of Equations (\ref{hpartial}), (\ref{hDeltahU}), (\ref{hdeltahbomega}), (\ref{wht}), (\ref{hbOmega2}), (\ref{hm}) and~(\ref{htPsi}),  we discover that
\begin{multline}
\hDelta_{\bR^\d}\hU+\frac{\partial}{\partial{}\wht}\,\hdelta\hbomega+\half\Vert\hbOmega\Vert^2-4\pi{}G\,\hm\,\vert\htPsi\vert^2
=
\lambda^{-2}\nu^{-2}
\Big[
\Delta_{\bR^\d}U+\frac{\partial}{\partial{}t}\delta\bomega
+\half\Vert\bOmega\Vert^2\\[4pt]
-4\pi{}G\,m\,\vert\tPsi\vert^2\,\nu^3\lambda^{2-\frac{\d}{2}}
\Big].
\label{hNC2}
\end{multline}
Equations (\ref{hNC1}) and (\ref{hNC2}) finally enable us to confirm that the NC field equations (\ref{NewtonCoriolis}) are indeed invariant under those chronoprojective transformations verifying $\lambda^{2-\frac{\d}{2}}\,\nu^3=1$, as already revealed in~(\ref{dilationConstraint}).

\subsection{The quantum representation of the SN group}

Let us finish this study by working out explicitly the quantum representation of the SN group in the cases $\d\neq4$ and $\d=4$ respectively.

\goodbreak

We claim that the mapping 
\begin{equation}
\rho(\Phi):\Psi\mapsto\Phi_*\Psi
\label{rho}
\end{equation}
provides a \textit{unitary} representation, $\rho$, of $\rSN(M,\rg,\xi)$ on the \textit{set of solutions, $\Psi$, of the \SN equations} (\ref{Yamabe}), (\ref{Mass}) and (\ref{NewtonSch}). Indeed, the push-forward mapping, $\Psi\mapsto\Phi_*\Psi$, is a group homomorphism, hence $\rho$ is a unitary representation in view of~(\ref{Unitarity}).

Let us work out the explicit form of the previously defined representation for the \SN group, $\rSN(\bR^{\d+2},\rg,\xi)$, of the spatially flat Bargmann structure~(\ref{BargmannQuater}).

\goodbreak

\bigskip
$\bullet$ We start with the generic case $\d\neq4$. The group to represent is defined by Equations (\ref{hbx}), (\ref{wht}) and (\ref{hs}) with $f=0$ (see (\ref{f=0})) and $d$ \& $g$ as in (\ref{dg}). In order to implement (\ref{rho}), i.e. $\rho(\Phi)\Psi=(\Phi^{-1})^*\Psi$, we need to compute $(\starbx,\start,\stars)=\Phi^{-1}(\bx,t,s)$. We readily find
\begin{eqnarray}
\label{starbx}
\starbx&=&A^{-1}\left[g\,\bx-\frac{(g\,t-e)}{d}\,\bb-\bc\right]\\[4pt]
\label{start}
\start&=&\frac{g\,t-e}{d}\\[4pt]
\label{stars}
\stars&=&
\nu\,s+g\la\bb,\bx\ra-\frac{g}{2d}\Vert\bb\Vert^2t
+\frac{e}{2d}\Vert\bb\Vert^2-\la\bb,\bc\ra-h
\end{eqnarray}
where, again, $A\in\rO(\d)$, $(\bb,\bc)\in\bR^\d\times\bR^\d$, $(e,h)\in\bR^2$, and $\nu\in\bR^*_+$, with $d$ and $g$ given by (\ref{dg}).
 
Moreover, referring to Equations (\ref{Psi(bx,t,s)}) and (\ref{hm}), we have
\begin{eqnarray}
\Phi_*\Psi
\nonumber
&=&
\Phi_*\left(e^{ims/\hbar}\,\psi(\bx,t)\,\vert\Vol(\rg)\vert^{\frac{\d}{2\d+4}}\right)\\[6pt]
\nonumber
&=&
e^{i(m/\nu)\stars/\hbar}\,(\pi(\Phi)_*\psi)(\bx,t)\,\vert\Vol(\Phi_*\rg)\vert^{\frac{\d}{2\d+4}}\\[6pt]
&=&
e^{i(m/\nu)\stars/\hbar}\,\psi(\starbx,\start)\,\lambda^{-\frac{\d}{4}}\vert\Vol(\rg)\vert^{\frac{\d}{2\d+4}}
\end{eqnarray}
and, using (\ref{lambdanuTer}) together with a slight abuse of notation, we obtain
\begin{equation}
\bigbox{
\begin{array}{rll}
\left[\rho(\Phi)\psi\right](\bx,t)
&=&
g^{\frac{\d}{2}}\,
e^{\frac{im}{\nu\hbar}\left(g\la\bb,\bx\ra-\frac{g}{2d}\Vert\bb\Vert^2t
+\frac{e}{2d}\Vert\bb\Vert^2-\la\bb,\bc\ra-h\right)}\,\times\\[5pt]
&&\psi\left(A^{-1}\left[g\,\bx-d^{-1}(g\,t-e)\,\bb-\bc\right],d^{-1}(g\,t-e)\right)
\end{array}
}
\label{rhodneq4}
\end{equation}
for all $\Phi=(A,\bb,\bc,e,h,\nu)\in\rSN(\bR^{\d+2},\rg,\xi)$, with $d$ and $g$ as in (\ref{dg}).

\goodbreak

For the subgroup of dilations generated by $\nu\in\bR^*_+$, we find
\begin{equation}
\medbox{
\left[\rho(\nu)\right]\psi(\bx,t)=\nu^{-\frac{3\d}{2(\d-4)}}\,\psi\big(\nu^{-\frac{3}{\d-4}}\bx,\nu^{-\frac{\d+2}{\d-4}}t\big)
}
\label{rhonu}
\end{equation}
which reduces, in the special case $\d=3$, to $\left[\rho(\nu)\psi\right](\bx,t)=\nu^{\frac{9}{2}}\,\psi\left(\nu^3\bx,\nu^5t\right)
$, in total accordance with the conclusion of~\cite{GG}.

\bigskip

$\bullet$ Let us consider now the special case $\d=4$, characterized by $\nu=1$ in (\ref{hbx}), (\ref{wht}) and (\ref{hs}). We first find the useful expression of $(\starbx,\start,\stars)=\Phi^{-1}(\bx,t,s)$, namely 
\begin{eqnarray}
\label{starbxBis}
\starbx&=&A^{-1}\left[\frac{\bx-\bb(g t-e)}{-f t+d}-\bc\,\right]\\[4pt]
\label{starttBis}
\start&=&\frac{g\,t-e}{-f t+d}\\[4pt]
\label{starsBis}
\stars&=&
s-\frac{f}{2}\frac{\Vert\bx\Vert^2}{-f t+d}
+\frac{\la\bb,\bx-\bb(g t-e)\ra}{-f t+d}
-\la\bb,\bc\ra
+\half\Vert\bb\Vert^2\frac{g t-e}{-f t+d}
-h
\end{eqnarray}
where, again, $A\in\rO(4)$, $(\bb,\bc)\in\bR^4\times\bR^4$, $(d,e,f,g,h)\in\bR^5$, and $dg-ef=1$.

\goodbreak

Skipping details, and with the same calculation as before, we end up with the following projective representation of the Schr\"odinger group, namely
\begin{equation}
\bigbox{
\begin{array}{rll}
\left[\rho(\Phi)\psi\right](\bx,t)
&=&
e^{
\frac{im}{\hbar}\left(
-\half\frac{f\Vert\bx\Vert^2}{-f t+d}
+\frac{\la\bb,\bx-\bb(g t-e)\ra}{-f t+d}-\la\bb,\bc\ra
+\half\Vert\bb\Vert^2\frac{g t-e}{-f t+d}
-h
\right)
}
\,\times
\\[6pt]
&&
\displaystyle
\frac{1}{(-f t+d)^{\frac{\d}{2}}}\,
\psi\left(
A^{-1}\!\left[\displaystyle
\frac{\bx-\bb(g t-e)}{-f t+d}-\bc
\right],
\displaystyle
\frac{g t-e}{-f t+d}
\right)
\end{array}
}
\label{rhod=4}
\end{equation}
for all $\Phi=(A,\bb,\bc,d,e,f,g,h)\in\Sch(\bR^{5,1})$.

\goodbreak

\section{Conclusion \& outlook}\label{ConclusionSection}

We have adopted, in this article, a geometric standpoint enabling us to propose, in terms of Bargmann structures over $(\d+1)$-dimensional Newton-Cartan structures, an intrinsic generalization of the \SN equation.

This allowed for a characterization of the maximal symmetry group of this generalized \SN equation, which we have named the ``\SN group''. The special case of spatially flat structures has been explicitly worked out for purpose of comparison with earlier work on the subject in dimension $\d=3$.
In doing so, we point out the special case $\d=4$ where the \SN group becomes isomorphic to the full Schr\"odinger group featuring, hence, inversions of the time-axis.

\goodbreak

It should be, however, noticed that the presence of the cosmological constant, $\Lambda$, in the NC field equations (\ref{NewtonFieldEqs}), viz., $\Ric(\rg)=(4\pi G\varrho+\Lambda)\,\theta\otimes\theta$ with $\varrho$ as in (\ref{NewtonSch}), \textit{breaks} the \SN group down to the Bargmann group. Indeed one should have $\lambda^{2-\frac{\d}{2}}\nu^3=\lambda^2\nu^2=1$ \& $\d>0$, hence $\lambda=\nu=1$.

Let us mention that the maximal symmetries of the L\'evy-Leblond-Newton equation can be uncovered using the same geometrical techniques; this is indeed the subject of a companion paper in preparation.

Finally, we would like to underline that there exists a notion of \textit{generalized symmetry} of the Schr\"odinger equation associated with the infinite-dimensional \textit{Schr\"odinger-Virasoro} group \cite{Hen,UR}. This symmetry is being fully geometrized in~\cite{DM} where the \SN equation is studied from this perspective; the brand new ``Schr\"odinger-Virasoro-Newton'' group is hence naturally introduced as the subgroup of the extended Schr\"odinger-Virasoro group for which the relations (\ref{S=0}) and (\ref{dilationConstraint}) hold. This work in progress is an extension of \cite{Gib} and highlights, in particular, the r\^ole of the Schr\"odinger-Virasoro-Newton group in the notable article~\cite{RT} about the Lie point symmetries of the \SN equation~(\ref{SNEq}).

\section*{Acknowledgments}
It is a pleasure to thank G. Gibbons for his interest in this work and for drawing our attention to key references, and also P. Horv\'athy for extremely useful suggestions. Special thanks are due to J.-P. Michel for a very careful and critical reading of the manuscript, and for most enlightening discussions.
This work has been carried out in the framework of the Labex ARCHIMEDE (Grant No. ANR-11-LABX-0033)
and of the A*MIDEX project (Grant No. ANR-11-IDEX-0001-02), funded by the ``Investissements d'Avenir''
French Government program managed by the French National Research Agency~(ANR).


\appendix

\section{Appendix}
\label{AppendixB}

\subsection{Conformally equivariant quantization}
\label{CEQ}

It has been proved in \cite{DO,DLO} that the Yamabe operator, see (\ref{Yamabe}), can be obtained via Conformally Equivariant Quantization (CEQ) of the quadratic Hamiltonian function
\begin{equation}
H=\rg^{\alpha\beta}(x)p_\alpha{}p_\beta
\label{H}
\end{equation}
on the cotangent bundle, $T^*M$, of a conformally flat pseudo-Riemannian manifold $(M,\rg)$ of signature $(n_+,n_-)$, with $\tn=n_++n_-$. This quantization scheme establishes, for generic values of weights $w,w'$, a unique iso\-morphism, $\cQ_{w,w'}$, of $\rO(n_++1,n_-+1)$-modules between the space of fiberwise polynomial functions of $T^*M$ and that of differential operators sending $w$-densities to $w'$-densities of $M$. For some ``resonant'' values of $w,w'$, uniqueness of CEQ is no longer guaranteed. However, imposing that the resulting differential operators be self-adjoint may restore uniqueness of CEQ. For example, in the case of the resonant values $w,w'=1-w$ with $w$ as in (\ref{w}), we obtain the self-adjoint Yamabe operator
\begin{equation}
\cQ_{\frac{\tn-2}{2\tn},\frac{\tn+2}{2\tn}}(H)=-\hbar^2\Delta_Y(\rg)
\label{QH}
\end{equation}
which we straightaway extend to any pseudo-Riemannian manifold.

\goodbreak

For Bargmann structures, the fundamental vector field $\xi$ (of the $(\bR,+)$-principal bundle $\pi:M\to{}\cM$) gives rise to a special Hamiltonian function on $T^*M$ endowed with its canonical $1$-form $\varpi$. Let us denote by $\xi^\sharp$ its canonical lift to~$T^*M$, so that $L_{\xi^\sharp}\varpi=(d\varpi)(\xi^\sharp)+dm=0$, where $m=\varpi(\xi^\sharp)$, i.e.,
\begin{equation}
m=p_\alpha\xi^\alpha
\label{m}
\end{equation}
is the \textit{momentum mapping} of the Hamiltonian $(\bR,+)$-action on the symplectic manifold $(T^*M,d\varpi)$. The function (\ref{m}) is interpreted as the \textit{mass} of the classical system~\cite{DBKP}. Then, CEQ applied to first-order fiberwise polynomials readily leads, in our case, to 
\begin{equation}
\cQ_{\frac{\tn-2}{2\tn},\frac{\tn-2}{2\tn}}(m)=\frac{\hbar}{i}\,L_\xi.
\label{Qm}
\end{equation}

\goodbreak

\subsection{Mass dilation}
\label{MassDilation}

Under a rescaling $\xi\mapsto\hxi=\nu\xi$, the mass (\ref{m}) thus transforms as
\begin{equation}
\hm=\nu\,m.
\label{hmTer}
\end{equation}

Consider now a conformal Bargmann diffeomorphism $\Phi$ (verifying (\ref{Conf}) and (\ref{AutFib})) and its canonical lift, $\Phi^\sharp$, to $T^*M$. Straightforward calculation yields then $(\Phi^\sharp)^*m
=(\Phi^\sharp)^*(\alpha(\xi^\sharp))
=\alpha((\Phi^\sharp)^*(\xi^\sharp))
=\alpha((\Phi^*\xi)^\sharp))
=\alpha(\nu\xi^\sharp)$, 
hence
\begin{equation}
(\Phi^\sharp)^*m=\nu\,m.
\label{PhiStarm}
\end{equation}
Again, this shows that the mass is rigidly dilated by conformal Bargmann transformations.

\newpage

\addcontentsline{toc}{section}{References}




\begin{thebibliography}{99}

\bibitem{BGDB}
M. Bahrami, A. Grossardt, S. Donati and Angelo Bassi,
``The Schr\"odinger equation and its foundations'',
{\em New Journal of Physics} {\bf 16} (2014) 115007.

\bibitem{Bar}
V. Bargmann, 
``On unitary ray representations of continuous groups'', 
{\em Ann. Math.} {\bf 59}:2 (1954) 1--46
.

\bibitem{Bes}
A. L. Besse, 
{\em Einstein manifolds},
Springer-Verlag (Berlin-Heidelberg 1987), first reprint~(2002).

\bibitem{Bri}
M. W. Brinkmann, 
``On Riemann spaces conformal to Euclidean spaces'',
{\em Proc. Natl. Acad. Sci. U.S.} {\bf 9} (1923) 1--3;
``Einstein spaces which are mapped conformally on each other'',
{\em Math. Ann.} {\bf 94} (1925) 119--145 .

\bibitem{BDP1}
G.~Burdet, C. Duval  and M.~Perrin,
``Cartan Structures on Galilean Manifolds: the Chronoprojective Geometry'',
{\em J. Math. Phys.} {\bf 24} (1983) 1752--1760.

\bibitem{BDP2}
G.~Burdet, C. Duval  and M.~Perrin,
``Chronoprojective Cartan Structures on Four-Dimensional Manifolds'',
{\em Publ. RIMS, Kyoto Univ.} {\bf 19} (1983) 813--840.

\bibitem{BDP3}
G.~Burdet, C. Duval  and M.~Perrin,
``Time-dependent Quantum Systems and Chronoprojective Geometry'',
{\em Lett. Math. Phys.} {\bf 10} (1985) 255--262.

\bibitem{Cari}
M. Cariglia,
``Hidden Symmetries of Dynamics in Classical and Quantum Physics'',
{\em Rev. Mod. Phys.} {\bf 86} (2014) 1283, 51 pp.

\bibitem{CGvHHZ}
M. Cariglia, G. Gibbons, J.-W. van Holten, P. Horv\'athy and P. M. Zhang,
``Conformal Killing tensors and covariant Hamiltonian dynamics'',
{\em J. Math. Phys.} {\bf 55} (2014) 122702, 32 pp.

\bibitem{Car}
E.~Cartan,
``Sur les vari\'et\'es \`a connexion affine et la
th\'eorie de la relati\-vit\'e g\'en\'eralis\'ee,''
{\em Ann. Sci. Ecole Norm. Sup.} (4) {\bf 40} (1923) 325--412.

\bibitem{Dio}
L. Di\'osi,
``Gravitation and quantum-mechanical localization of macro-objects'',
{\em Phys. Lett.} {\bf A 105} (1984) 199--202.

\bibitem{Duv1}
C. Duval,
``Quelques proc\'edures g\'eo\-m\'etri\-ques en dynamique des particules'', 
Doctorat d'Etat \`es Sciences (Aix-Marseille-II), 1982 (unpublished).

\bibitem{Duv2}
C. Duval,
``Sur la g\'eom\'etrie chronoprojective de l'espace-temps classique'',
{\em in Actes des Journ\'ees relativistes de Lyon},
(1982). 

\bibitem{DBKP}
C. Duval, G.~Burdet, H. P.~K\"unzle and M.~Perrin,
``Bargmann structures and Newton-Cartan theory'',
{\em Phys. Rev.} {\bf D 31} (1985) 1841--1853.

\bibitem{Duv3}
C. Duval,
``Nonrelativistic Conformal Symmetries and Bargmann Structures,''
in {\em Proc. Symposium Conformal Groups and Related Symmetries.
Physical Results and Mathematical Background},
Clausthal 1985, 
(A.O.~Barut \& H.D.~Doebner Eds),
pp 162--182,
Lecture Notes in Physics 261,
Springer-Verlag (1986). 


\bibitem{Duv4}
C. Duval,
``The Dirac \& L\'evy-Leblond Equations and Geometric Quantization'', 
in {\em Proc. XIV International Conference on Differential Geometric Methods in
Mathematical Physics}, 
(P.L.~Garcia \& A.~P\'erez-Rendon Eds),
Salamanca 1985,
pp 205--221,
Lecture Notes in Mathematics {\bf 1251},
Springer-Verlag (1987).

\bibitem{DGH}
C. Duval, G.~Gibbons and P.~Horv\'athy,
``Celestial mechanics, conformal structures, and gravitational waves'',
{\em Phys. Rev.} {\bf D 43} (1991) 3907--3922.

\bibitem{DH}
C. Duval and P.~Horv\'athy,
``Non-relativistic conformal symmetries and Newton-Cartan structures'',
{\em J. Phys. A: Math. Theor.} {\bf 42} (2009) 465206 (32pp).

\bibitem{DHP1}
C. Duval, P.~Horv\'athy and L.~Palla,
``Conformal symmetry of the coupled Chern-Simons and gauged nonlinear
Schr\"odinger equations'',
{\em Phys. Lett.} {\bf B 325} (1994) 39--44.



\bibitem{DHP4}
C. Duval, P.~Horv\'athy and L.~Palla,
``Spinors in Non-Relativistic Chern-Simons Electro\-dynamics'', 
{\em Annals Phys.}, \textbf{249} (1996) 265--297.

\bibitem{DL}
C. Duval and S. Lazzarini,
``Schr\"odinger manifolds'',
{\em J. Phys. A: Math. Theor.} {\bf 45} (2012) 395203 (24 pp).


\bibitem{DM}
C. Duval and J.-P. Michel,
``Geometric realization of the Schr\"odinger-Virasoro group'',
in preparation.

\bibitem{DLO}
C. Duval, P. Lecomte and V. Ovsienko,
``Conformally Equivariant Quantization : Existence and Uniqueness'',
{\em Ann.~Inst. Fourier} \textbf{49}:~6 (1999) 1999--2029.

\bibitem{DO}
C. Duval and V. Ovsienko,
``Conformally Equivariant Quantum Hamiltonians'',
{\em Selecta Math. New Ser.} \textbf{7} (2001) 291--320.

\bibitem{EK}
J. Ehlers, and W. Kundt,
``Exact solutions of the gravitational field equations'',
in \textsl{Gravitation: An introduction to current research},
L. Witten Ed., Wiley, New York, 1962.

\bibitem{Eis}
L. P. Eisenhart,
``Dynamical trajectories and geodesics'',
{\em Ann. of Math.} {\bf 30}(1929) 591--606.

\bibitem{Gib}
G. Gibbons,
``Dark Energy and the Schwarzian Derivative'',
\texttt{arXiv:1403.5431 [hep-th]}.

\bibitem{GG}
D. Giulini and A. Grossardt,
``Gravitationally induced inhibitions of dispersion according to the Schr\"odinger-Newton equation'', 
{\em Class. Quantum Grav.} \textbf{28} (2011) 195026 (17pp).

\bibitem{GG2}
D. Giulini and A. Grossardt,
``Gravitationally induced inhibitions of dispersion according to the Schr\"odinger-Newton equation for a homogeneous-sphere potential'', 
{\em Class. Quantum Grav.} \textbf{30} (2013) 155018.

\bibitem{GP}
 J. Gomis, and J.M. Pons,
``Poincar\'{e} Transformations and Galilei Transformations'',
{\em Phys. Lett.} {\bf A 66} (1978), 463--465.


\bibitem{Hag}
C. R. Hagen,
``Scale and conformal transformations in Galilean-covariant field theory,''
{\em Phys. Rev.} {\bf D 5} (1972) 377--388.

\bibitem{Hen} 
M. Henkel 
``Schr\"odinger invariance and strongly anisotropic critical systems,''
{\em J. Stat. Phys.} {\bf 75} (1994) 1023--1061.

\bibitem{Jac} 
R. Jackiw, 
``Introducing scaling symmetry'',
{\em Phys. Today} {\bf 25} (1972) 23--27.

\bibitem{JP}
R. Jackiw and S.-Y. Pi,
``Soliton solutions to the gauged nonlinear Schr\"odinger equation on the plane,''
{\em Phys. Rev. Lett.} {\bf 64} (1990) 2969--2972;
`Classical and quantal non\-relativistic Chern-Simons theory'',
{\em Phys. Rev.} {\bf D 42} (1990) 3500--3513.

\bibitem{Kun}
H. P. K\"unzle,
``Galilei and Lorentz structures on space\-time:
Comparison of the corresponding geometry and physics'',
{\em Ann. Inst. H. Poincar\'e, Phys. Th\'eor.}
{\bf 17} (1972) 337--362, and references therein.

\bibitem{MRS}
J.-P. Michel, F. Radoux and J. Silhan,
``Second Order Symmetries of the Conformal Laplacian'',
{\em SIGMA} {\bf 10} (2014) 016, 26 pages.

\bibitem{MPT}
I. M. Moroz, R. Penrose and P. Tod,
``Spherically-symmetric solutions of the Schr\"odinger-Newton equations'',
{\em Class. Quantum Grav.} {\bf 15} (1998) 2733--2742.

\bibitem{Nie} 
U. Niederer,
``The maximal kinematical symmetry group of the free Schr\"odinger equation'',
{\em Helv. Phys. Acta} {\bf 45} (1972) 802--810.  

\bibitem{Pen1}
R. Penrose,
``On Gravity's Role in Quantum State Reduction'',
{\em Gen. Relat. Grav.} {\bf 28} (1996) 581--600.

\bibitem{Pen2}
R. Penrose,
``On the Gravitization of Quantum Mechanics I : Quantum State Reduction'',
{\em Found. Phys.} (2014) 44 : 547--575.

\bibitem{RT}
O. Robertshaw and P. Tod,
``Lie point symmetries and the geodesic approximation of the Schr\"odinger-Newton equations'', 
{\em Nonlinearity} {\bf 19} (2006) 1507--1514.

\bibitem{JMS}
J.-M. Souriau, 
``Grammaire de la Nature'',
\url{http://www.jmsouriau.com/Grammaire_de_la_nature.htm}.

\bibitem{TM}
P. Tod and I. M. Moroz,
``An analytical approach to the Schr\"odinger-Newton equations'',
{\em Nonlinearity} {\bf 12} (1999) 201--216.

\bibitem{Tra}
A.~Trautman,
``Sur la th\'eorie newtonienne de la gravitation,''
{\em C.R. Acad. Sci. Paris} {\bf 257} (1963) 617;
``Comparison of Newtonian and relativistic theories of
space time,'' pp.~413--425 in  {\sl Perspectives in Geometry and
Relativity},  (B.~Hoffmann, Ed.),
Indiana University Press, Bloomington, 1964.

\bibitem{UR}
J. Unterberger and C. Roger,
{\em The Schr\"odinger-Virasoro Algebra: Mathematical structure and dynamical Schr\"odinger symmetry},
Springer-Verlag, Berlin Heidelberg (2012).


\end{thebibliography}
\end{document}